\documentclass[12pt]{iopart} 

\usepackage{latexsym} 
\usepackage{epsf}
\usepackage{times}   

\def\footnoterule{\noindent\hbox to 6em{\hrulefill}\vskip6pt}

\begin{document}  

\title[]{Gaps between equations and experiments in\\
quantum cryptography}

\author{John M. Myers} 

\address{Gordon McKay Laboratory, Division of Engineering and Applied
Sciences,\\ Harvard University, Cambridge, MA 02138\\}

\author{and\\}

\author{F. Hadi Madjid}

\address{82 Powers Road, Concord, MA 01742}   
 
\maketitle 

\begin{abstract}

Traditional methods of cryptographic key distribution rest on judgments
about an attacker.  With the advent of quantum key distribution (QKD)
came proofs of security for the mathematical models that define the
protocols BB84 and B92; however, applying such proofs to actual
transmitting and receiving devices has been questioned.

Proofs of QKD security are propositions about models written in the
mathematical language of quantum mechanics, and the issue is the linking
of such models to actual devices in an experiment on security. To explore
this issue, we adapt Wittgenstein's method of {\em language games} to
view quantum language in its application to experimental activity
involving transmitting and receiving devices.

We sketch concepts with which to think about models in relation to
experiments, without assuming the experiments accord with any model;
included is a concept of one quantum-mechanical model {\em enveloping}
another. For any model that agrees with given experimental results and
implies the security of a key, there is an enveloping model that agrees
with the same results while denying that security.  As a result there is
a gap between equations and the behavior recorded from devices in an
experiment, a gap bridged only by resort to something beyond the reach of
logic and measured data, well named by the word {\em guesswork}.

While this recognition of guesswork encourages eavesdropping, a related
recognition of guesswork in the design of feedback loops can help a
transmitter and receiver to reduce their vulnerability to eavesdropping.

\end{abstract} 
 
\vfil
\eject
\section{Introduction}\label{sec:1}

In a laboratory that has equations on a blackboard and
devices\footnote[1]{We think of `devices' as being designed using quantum
mechanics to serve practical purposes, and of `instruments' as being used
to test models in quantum mechanics.  Often the same piece of gear can
play either role, and we shall use the word `device' to mean either.} on
a bench, a fair question is how do the equations and the devices
connect?  The terms of the equations act as words of a language used in
the laboratory.  People say or write these terms along with
non-mathematical words ({\em e.g.}\ `screwdriver') while they work with
devices, an activity Wittgenstein characterized as a {\em language game}
\cite{w1,w2,w3,w4}.  Recently we brought the writing of equations and the
using of devices into a common language game.  This allowed us to prove
that deciding to describe a particular laboratory situation by a
particular quantum mechanical model takes something beyond logic and
measured data, something aptly called a {\em guess} \cite{ams}.

Here are three examples of questions in quantum cryptography resolvable
only by augmenting equations and data by guesswork:

1. Alice and Bob interpret some equations as describing a transmitter and
a receiver for quantum-cryptographic key distribution.  How do they make
or buy devices that behave in accord with those equations?

2. Bob uses quantum-electrodynamics to define photons in terms of
creation operators, and Bob's photodiode detector feeds current blips
into an audio amplifier so he hears and counts them. Under what
circumstances does counting blips give him a claim to say he counts
photons?

3. Alice and Bob experiment with a transmitter and a receiver linked by
an optical fiber, and their results agree with the predictions of a model
in which it is proved they are secure against undetected eavesdropping
using individual attacks.  What use can they make of that proof?

We will show that quantum states are by no means `plainly visible' in the
devices of experiments on quantum cryptography; rather there is a gap, as
pictured in figure 1, between states as terms in equations and results
recorded from devices on a laboratory bench.  To show this gap and its
bridging by guesswork, we will sketch a conceptual framework in section
2, and then turn in section 3 to quantum cryptographic key distribution
as an appealing arena for examples.  Cryptographers know that relying on
a theory does not bind one's enemy to rely on the same theory.  They know
too that abstractions endemic to theories are likely targets for
cryptographic attack, so they can well imagine that a proof of the
security of quantum key distribution might not be the last word
\cite{lowry}.  By seeing gaps in physics where logic cannot reach and
guesses are indispensable, we will see both limitations to proofs of the
security of quantum key distribution and opportunities to improve
security.

\section{Concepts for bridging equations and experiments}\label{sec:2}

Preliminary to showing how guesses bridge the gap between equations and
experiments, we introduce the concept of a quantum mechanical model of an
experimental trial; we also attend to {\em trial-to-trial management} and
introduce the {\em envelopment of one model by another} as a building
block for sections \ref{sec:3} and \ref{sec:4}.\looseness=-1

\subsection{Models of experimental trials}\label{subsec:2.1}

Quantum mechanics provides language in which to write equations for
various probabilities and to relate these equations logically to one
another.  Proofs of QKD security apply equations for probabilities as
models of devices expected to be used to communicate keys.  Quantum
mechanical operators and state vectors are germane to QKD security only
as they appear as components in equations that define probabilities in
models that someone applies to actual or anticipated devices.

To test a model of QKD one subjects the devices to a run of experimental
trials, each of which involves setting up devices and recording results. 
In order that trials of devices can test a model, the model must speak of
the devices, and for this the operators and state vectors must somehow be
linked to these devices.

Appreciating the need to link state vectors to devices requires you to
resist a temptation.  In planning a test of a model that involves quantum
states, avoid picturing yourself `in a Hilbert space of these states,' an
intangible world disconnected from the world of prisms and fibers on a
bench.  Instead, look at how you can use quantum states in calculations. 
Momentarily set aside the picture of state vectors as denizens of a
Hilbert-space world, and notice how you write characters for quantum
states into a computer memory, preferably the memory of the same computer
that manages the devices during a run of experimental trials.  This puts
your writing of quantum states and calculating with them into a common
world with the handling of prisms and fibers.

That still leaves the problem of how to attach state vectors and
operators, as terms in equations of a model, to devices in a laboratory. 
One goes half way by deciding to attach states and operators as terms in
equations to various settings of the knobs and levers by which the
devices are configured in a trial, to get state-valued functions and
operator-valued functions of the knob and lever settings \cite{ams}.  To
complete the attachment, one has to decide which device settings go with
which states and operators.  Once accomplished, the attachment of the
states and operators to device settings converts the usual quantum
mechanical probabilities of outcomes for given states and operators to
probabilities of outcomes for given device settings.  With this trick one
gets equations interpreted as asserting something about devices.  The
first kind of model to consider can be called a quantum-mechanical model
of devices or, for short, QM-model.  Later we will use models of this
type as a foundation on which to build some other kinds of models.

A QM-model $\alpha$ consists of a triple of functions that, in
combination, generate probabilities of outcomes for various set-ups of
devices.  Simplifying the full description given in appendix A, one can
view these functions as follows.  (1) State preparation is expressed by a
function $|\mbox{state}_\alpha (\mbox{set-up})\rangle$ from the domain
of set-ups of devices to state vectors; (2) evolution over a time
duration is expressed by a function $U_\alpha (\mbox{set-up})$ from the
domain of set-ups of devices to unitary operators; and (3) a measurement
process that generates results is expressed by a function from set-ups of
devices to operators, most generally in the form of a positive
operator-valued measure (POVM) as a set of detection operators, one for
each possible outcome, $M_{\alpha,{\rm outcome}}(\mbox{set-up})$
\cite{helstrom}. This triple of functions generates probabilities of
outcomes as a function of set-up, asserted by model $\alpha$, according
to the usual rule:
\begin{eqnarray}
\lefteqn{\Pr\nolimits_{\alpha}(\mbox{outcome}|\mbox{set-up})}\nonumber\\
&=\langle \mbox{state}_\alpha \mbox{(set-up)}|U_\alpha(\mbox{set-up})
M_{\alpha,{\rm outcome}}(\mbox{set-up})
U^{\dag}_\alpha(\mbox{set-up})| \mbox{state}_\alpha \mbox{(set-up)}
\rangle ,\nonumber\\ \end{eqnarray} where the subscripts $\alpha$
indicate that the functions to which they are attached are particular
to model $\alpha$ so another model would involve other functions.

Analyzing the receipt of a cryptographic key from the standpoint of
quantum mechanics, one is interested in the question the other way
around. Given an outcome, Bob (or Eve) wants to know what state Alice
transmitted, a subject addressed by quantum decision theory via
Bayes rule \cite{helstrom,holevo}.

\subsection{Computer management of trials and models}\label{subsec:2.3}

In order to describe the set-up of devices sufficiently to prove the need
for guesswork in choosing a model for QKD, we restrict ourselves to
describing only a part of setting up that can be controlled by a
Classical digital Process-control Computer (CPC).  This part is described
by the commands transmitted from the CPC to actuators internal to these
devices.  Such commands call, for example, for a stepping motor to rotate
a polarizer to a certain angle.  The same CPC serves to record the
writing of QM-models in which set-ups are expressed as CPC-commands and
to execute quantum-mechanical calculations involving these models.  (The
proof of Proposition~1 in section \ref{sec:3} assumes only that some
choices of set-up can be CPC mediated, not that everything in setting up
an experiment is CPC mediated.)  As pictured in figure 2, the CPC:

(1) provides an investigator with a display by which to experience
results of trials reported by the devices to the computer; 

(2) gives the investigator a keyboard by which to operate the computer; 

(3) makes a record of what the investigator does (such as entering
equations into programs that, when executed, control the devices); and 

(4) (in various ways) allows feedback from the devices to modify the
equations \cite{ams}.

Section \ref{sec:3} presents examples in which devices are controlled
jointly by Alice, Bob, and Eve, each via a CPC; in this case a `command'
is taken to be the concatenation of commands from the three CPCs of
Alice, Bob, and Eve, respectively.

\subsection{Envelopment of one QM-model by another}\label{subsec:2.4}  

In testing a QM-model against results of experimental trials, one tests
the probability distribution calculated from the model against relative
frequencies extracted from measured results.\footnote[2]{As discussed in
\cite{ams}, that extraction itself requires guesswork.} Any such test of
the probability distribution tests the triple of functions that define
the model, but only in the combination of equation (1); one cannot test any
of the functions separately without assuming the others. In particular,
one cannot test quantum states independently of the other two
functions.\footnote[3]{Claims that quantum tomography allows a scientist
to determine a state experimentally \cite{tomog1,tomog2} assume that the
scientist knows the operators that describe the measuring instrument. 
But how does the scientist determine these operators?  By using the
instruments to `measure known states': without guesswork, the scientist
is stuck in a circle, as an example to come will show.}  When the
probabilities of a model $\alpha$ are found to closely match relative
frequencies of outcomes for the various settings used in a run of
experimental trials of devices, one can still ask what other models agree
as well with the same relative frequencies.

To sharpen this question we define the first of two relations among
QM-models. By restricting the set of commands of a QM-model to a subset,
one generates what can be called a restriction of the QM-model.  When an
experiment confirms a QM-model, it actually confirms, at most, the
restriction of the QM-model to commands tested, and often these are a
small subset of the commands of the model. Suppose model $\alpha$ is such
a restriction, and that it generates probabilities that closely match the
relative frequencies of a run of trials.  Then any model $\beta$ agrees
with the experimental trials if and only if it agrees with this model
$\alpha$.  Hence we want to know: what models other than model $\alpha$
predict the same probabilities as does $\alpha$ concerning the settings
covered by $C_\alpha$?  (Notice that this question is made possible by
the attachment of states and operators to settings of devices.) As we
shall see, for any QM-model $\alpha$ a large set of other QM-models,
defined by a variety of triples of functions, produce the same
probabilities for settings of $C_\alpha$.  As a result, a tight screening
of probability distributions by experiments induces only a loose
screening of state- and operator-valued functions.  This looseness in the
testability of quantum states limits the role of proofs in quantum
cryptography.

To clarify this looseness we introduce the concept of an {\em
envelopment} of one model by another.  Let $O_\alpha$ be the set of
outcomes for model $\alpha$. Given two QM-models $\alpha$ and $\beta$,
suppose there is a function $f$ from a subset of $C_\beta \times O_\beta$
to $C_\alpha \times O_\alpha$, with the property of preserving
probabilities in the sense that: 
\begin{equation} (\forall\
b \in C_\alpha, j \in O_\alpha)\ \Pr\nolimits_\alpha(j|b) =
\sum_{(b',j') \in f^{-1}(b,j)}
\Pr\nolimits_\beta(j'|b'). \end{equation} 
We say such a function $f$ {\em envelops} $\alpha$ by $\beta$.  An
enveloping model $\beta$ need not preserve inner products of quantum
states of model $\alpha$, and for this reason the inner products of
quantum states cannot be determined from experimental results without
augmentation by guesswork.  In the case to be discussed in section
\ref{sec:3}, there is no `mixing' by the mapping $f$, which is to say
there exist functions $g$ and $h$ such that $f(b,j)= (g(b),h(j))$, and the
envelopment of model $\alpha$ by model $\beta$ exhibits the provocative
inequality: 
\begin{equation} |\langle
\psi_\beta(b)|\psi_\beta(b') \rangle | \ll |\langle
\psi_\alpha(g(b))|\psi_\alpha(g(b')) \rangle| , 
\end{equation}
with implications for security shortly to be developed.

\section{Application to quantum cryptography}\label{sec:3}

For quantum key distribution using the protocols BB84 \cite{BB84} or B92
\cite{B92}, published arguments for the security of the key 
\cite{BB84,B92,ekert94,fuchs,slutskyPRA,slutskyAO,shor} assume a
QM-model $\alpha$ (augmented by Bayes rule), according to which, at each
trial, Alice's transmitter receives a command $b_A$ from some finite set
of possible commands and `prepares a unit state vector'
$|\psi_\alpha(b_A) \rangle$. In both early work that ignores noise and in
later work that accounts for noise (and invokes privacy amplification
\cite{slutskyPRA,slutskyAO,bennett95}), claims of security assume that
inner products of the `unit state vectors' that Alice can transmit are
experimentally verifiable without regard to what devices Eve might
invent.  The argument is that, given suitable magnitudes of inner
products $S_\alpha(b_A,b'_A) \equiv |\langle
\psi_\alpha(b_A)|\psi_\alpha(b'_A) \rangle |$, any snooping by Eve
disturbs Alice's states in ways Alice and Bob can almost surely detect as
an increase in the error rate in Bob's reception or the rate of
inconclusive outcomes \cite{brandt2} or both (when they compare notes
publicly, at the sacrifice of some of the bits of the key).

To what extent can experimental results on devices confirm a model
$\alpha$ that, based on inner products, asserts security of QKD? Alice
and Bob need to be secure not just against an attack by snooping devices
that they use when they play Eve's part in Red-Blue exercises, but
against all of the attacks, within some class, that a real Eve might
invent.  Are Alice's inner products verifiably independent of Eve's
inventions?

To claim experimental confirmation of the security implied by inner
products of model $\alpha$ is to claim that 
\begin{enumerate}
\item[a)] the probabilities calculated from model $\alpha$ match
closely the relative frequencies of experimental results, and
\item[b)] no other QM-model (say one with smaller inner products) that
challenges that security produces the same probabilities that match
those relative frequencies.\end{enumerate} 
Although attachments of models of quantum cryptography to actual devices
have been questioned \cite{slutskyPRA,lo}, the possibility of
enveloping a model $\alpha$ by a model $\beta$ having smaller inner
products went unnoticed, with the consequence that the need to make the
claim (b)---no model with smaller inner products fits the experimental
results---also went unnoticed.

The published arguments for security assume ``any attack that Eve might
invent'' is equivalent to ``any QM-model that Eve might implement that
has the inner products of model $\alpha$''.  Thus the experimental
confirmation of security depends on the notion that inner products are
experimentally testable.  The trouble is that, as shown in section
\ref{sec:2}, experiments cannot test states and their inner products {\em
per se} but only the probabilities they generate.  For example, suppose a
run of experimental trials shows a close match to the probabilities of a
QM-model $\alpha$ that speaks of certain inner products
$S_\alpha(b_A,b'_A)$.  Do these trials `demonstrate ({\em e.g.}
single-photon) states' having these inner products?  The answer is
``{\em no},'' because for any model $\alpha$ there is always an
envelopment by an alternative model $\beta$ that produces probabilities
in agreement with $\alpha$ concerning the experimental results that Alice
and Bob have on hand, but generates these probabilities on the basis of
different states with smaller inner products.  Because of its smaller
inner products, the model $\beta$ points to something outside of model
$\alpha$ that, if Eve can do it, allows her to eavesdrop undetected.

\subsection{Example of conflicting models that fit the same measured
data}\label{subsec:3.1}

Among widely used models of the security of quantum key distribution
against undetected individual attacks, there are two cases to consider,
corresponding to Eve measuring Alice's signal directly, or, using a
probe, indirectly.  Deferring models of Eve's use of a probe to appendix
B, we consider an individual eavesdropping attack in which Eve measures
Alice's signal and tries to transmit a signal to Bob, aiming to match
what would have come from Alice.  Any model $\alpha$ asserting security
against this attack rests on an inner product for Alice's possible
states.  We now show how to envelop model $\alpha$ by a model $\beta$
having a smaller inner product but producing the same probabilities as
model $\alpha$.  

The model $\alpha$ assumes that the command set $C_\alpha$ consists of
concatenations of a command $b_A$ from Alice to determine a state vector
and a command $b_E$ from Eve to select a POVM.  These commands produce
Alice's state vector $|v_\alpha(b_A) \rangle \in {\cal H}_\alpha$, and
Eve's measurement expressed by a POVM $M_\alpha(b_E)$ which has a
detection operator $M_\alpha(b_E;j_E)$ acting on ${\cal H}_\alpha$,
associated with outcome $j_E$.  Model $\alpha$ implies that the
conditional probability of Eve obtaining the outcome $j_E$ given her
command $b_E$ and Alice's command $b_A$ is \goodbreak
\begin{equation} \Pr\nolimits_\alpha(j_E|b_A,b_E) = \langle
v_\alpha(b_A) | M_\alpha(b_E;j_E)| v_\alpha(b_A) \rangle,
\label{eq:p_alpha}\end{equation} implying that the error rate for Eve in
distinguishing among Alice's commands, asserted by model $\alpha$,
depends on the inner products $S_\alpha(b_A|b'_A)$.

\vskip\abovedisplayskip
\noindent {\bf Proposition 1}\hskip1em {\em
 Given any such model $\alpha$ with inner products
$S_\alpha(b_A,b'_A)$ and given any $0 \leq r < 1$, there is a model
$\beta$ that gives the same conditional probabilities of Eve's
outcomes for all her commands belonging to $E_\alpha$, so that
\begin{equation} (\forall\ b_A \in A_\alpha, b_E
\in E_\alpha)\ 
\Pr\nolimits_\beta(j_E|b_A,b_E) = \Pr\nolimits_\alpha(j_E|b_A,b_E)
\label{eq:same} \end{equation} 
while 
\begin{equation}
S_\beta(b_A,b_A') \leq r S_\alpha(b_A,b'_A). \label{eq:inner}
\end{equation} }
\par\vskip\belowdisplayskip

\noindent {\em Proof}: Motivated by the idea that, unknown to Alice,
her transmitter signal might generate an additional ``leakage'' into
an unintended spurious channel that Eve reads, we construct the
following enveloping model $\beta$ which assumes: 
\begin{enumerate}
\item the same set of commands for Alice, \item a larger Hilbert space
${\cal H}_\beta = {\cal H}_{\rm leak} \otimes {\cal H}_\alpha$ in
which Alice produces vectors $|v_\beta(b_A) \rangle = |w_\beta(b_A)
\rangle \otimes |v_\alpha(b_A) \rangle$, with $|w_\beta(b_A) \rangle
\in {\cal H}_{\rm leak}$; \item a larger set of commands for Eve,
$E_\beta = E_\alpha \sqcup E_{\rm extra}$ (disjoint union); \item a
POVM-valued function of Eve's commands to her measuring instruments,
with detection operators 
\begin{equation} M_\beta(b_E;j_E) = \left\{
\begin{array}{l} {\bf 1}_{\rm leak} \otimes M_\alpha(b_E;j_E) \mbox{
for all } b_E \in E_\alpha,\\ \mbox{Eve's choice of POVM to
distinguish }  \\ \quad|v_\beta(0)\rangle\mbox{ from } |v_\beta(1)
\rangle
\mbox{ if } b_E \in E_{\rm extra}. \end{array} \right. 
\end{equation}
\end{enumerate} 
According to model $\beta$, if Eve chooses any measurement command of
$E_\alpha$, equation (\ref{eq:p_alpha}) holds.  But model $\beta$ speaks not
of the vectors $|v_\alpha(b_A) \rangle$ but of other vectors having an
inner product of magnitude
\begin{eqnarray} S_\beta(b_A,b'_A) &\stackrel{\rm def}{=}& |\langle
v_\beta(b_A)|v_\beta(b'_A) \rangle |\nonumber\\ 
 &= &|\langle
w(b_A)|w(b'_A)
\rangle | |\langle v_\alpha(b_A)|v_\alpha(b'_A) \rangle|.
\label{eq:winner}
\end{eqnarray} 
We can choose the unit vectors $|w(b_A) \rangle$ at will; in particular
nothing excludes choosing them so that $\forall\ b_A \neq b'_A$,
$\,|\langle w(b_A)|w(b'_A) \rangle | \leq r$, from which the proposition
follows.  $\Box$

\subsection{Impossibility of positive tests of inner
products}\label{subsec:3.2}

Models of the form of model $\beta$ with its $r < 1$ thus match any data
that match model $\alpha$ while denying security of the key. Given the
two QM-models, $\alpha$ and $\beta$, that disagree about cryptographic
security while agreeing about probabilities for commands of $C_\alpha$,
one would like to decide between the conflicting models by an
experiment.  But, without guesswork, this is impossible, for the
enveloping model $\beta$ is no description of instruments on hand;
instead, it is a picture of behavior that Eve might try to achieve by
inventing a snooping device not yet known.  Nor does model $\beta$ (nor
any QM-model) say of itself how to implement it, because it speaks only
of states and operators rather than of the prisms and optical fibers that
one can find in a supply room.  Implementing model $\beta$ would
require invention and discovery, such as gaining access to a channel
carrying leakage states \cite{brandt3,huttner}.  For this reason there
can be no positive experimental test of the inner products of quantum
states, nor of claims of cryptographic security based on such inner
products.  Indeed, to claim security against `all the eavesdropping
devices that Eve might invent' for individual attacks is to misuse the
word `all'; by confusing `all devices' described by a model $\alpha$ with
`all the models' consistent with experiments that accord with
probabilities expressed by model $\alpha$.  It is for this reason that
the application of a model to actual devices takes guesswork.  The notion
that experiments can determine Alice's inner products independent of
Eve's inventions is faulty, because it confuses experimental validation
of probabilities, which is possible, with experimental validation of
inner products of quantum states, which, as we have shown, is impossible
to disentangle from guesswork.

In more practical terms, to experimentally confirm QKD security by
`seeing a single-photon state,' one must see the absence of correlated
signals that accompany it, which Eve might receive, and nothing in Alice
and Bob's modeling and experiments can exclude a real Eve from finding
such a signal in a place Alice and Bob never thought to look, such as
light emitted by Alice's transmitter in a spurious frequency band
\cite{slutskyPRA}.
 
\section{Quantum physics between trials: a glimpse of
feedback}\label{sec:4} 

It is often advantageous to introduce feedback into a run of trials,
using the outcome of one trial in setting up a next trial, as 
illustrated in figure 3 \cite{jakeman,milburn,hj93,haus,shapiro,wiseman,%
doherty1,doherty2,doherty3}.\footnote[4]{We do not discuss `coherent
quantum feedback' in the sense of ref.\ \cite{lloyd}} By studying
trial-to-trial adjustment brought about by feedback loops designed using
QM-models, we will win fresh insight into the bridging of equations and
devices of experimental trials. The typographical view of equation
writing introduced in section \ref{sec:2}, with CPCs as equation
holders, allows one to view a feedback loop as a special sort of device,
a device containing an equation-holding CPC which a scientist uses to
link equations of QM-models (and other equations) to operations of other
devices.  This picture opens up a place in theoretical physics in which
to investigate trial-to-trial adjustment.

To define a feedback loop, a designer specifies a {\em control
function} that maps results of completed trials and other records on
hand to actions to be taken in response to these.  The specification
of the control function can make use of QM-models.  But neither a
QM-model nor a Schr\"odinger equation expresses a place in which an
outcome from one trial can enter to influence a state preparation at a
subsequent trial; on the contrary, trial-to-trial adjustments of
devices take place in a space outside of QM-models, so to speak.  To
deal with feedback in a quantum context, one has to implant QM-models
as components of larger models, which we call {\em control} models,
which contain control functions that express how a CPC responds to
outcomes.  Its need for feedback cements quantum physics into a
classical environment of CPCs that command the devices and record the
results.

\subsection{Comparisons of theory and experiment, in the presence of
feedback}\label{subsec:4.1}

In a feedback situation, CPCs that accompany other devices or are embedded
in them are part of the experimental instrumentation, and so are their
files housing the equations that define a control function.  These
files of equations are used not for analysis but to define actions of
devices, which puts an end to any categorical distinction between
equations and devices.  This poses a question of just what one compares
with what in an experimental trial, for one cannot, in this situation,
compare an `equation as something unphysical' against `behavior of
physical devices free of equations'.  One does, however, compare
predictions calculated from a QM-model and other equations of a
control model with experimental trials in which the same equations
take part in programs of CPCs that control the devices.  {\em I.e.},
one compares a `use of equations and guesses to make a prediction'
against `behavior of devices controlled by CPCs programmed using the
same equations' (and possibly other equations as well).

\subsection{Timing in quantum cryptography}\label{subsec:4.2} 

Quantum cryptography provides an arena for examples of trial-to-trial
adjustments, pertaining to timing.  In quantum key distribution, Bob's
receiver is intended to detect a sequence of signals transmitted by
Alice.  Bob's receiver (as well as Eve's snooping devices) depends on
keeping in step with the signals of the sequence transmitted by Alice.
Besides the need to correlate detections with Alice's acts of
transmission, there can be a need to gate a receiver off except in a
narrow time interval around the arrival of each signal.

Keeping the receiver in step with the signals arriving from the
transmitter requires one or another use of results of one trial to
regulate the receiver timing for another trial.  That is
synchronization, an important form of trial-to-trial adjustment.  For
understanding synchronization, it is necessary to look for a quantum
description of reception that describes a receiver only imperfectly
timed.  This leads us to the notion of a receiver that measures, or at
least estimates, the time difference between the arrival of a signal
and the gating of its detector, a time difference we call {\em skew}.
For a quantum-mechanical model to express this additional distinction
of temporal skew in detection, it has to allow for a larger variety of
quantum states than did models $\alpha$ and $\beta$ above, with the
skew as an additional argument, which entails expanding the dimension
of the Hilbert space beyond that of model $\alpha$.

To glimpse an expanded model $\gamma$ that envelops the model $\alpha$
of section \ref{sec:3}, and that allows a designer to picture timing,
suppose a receiver is designed to expect the $k$-th signal at reading
$t_k$ of its clock, and that the clock is fast by a {\em skew} $s_k$;
then the signal arrives not at clock reading $t_k$ but in a small interval
around $t_k - s_k$. To allow for skew, model $\gamma$ must have a
Hilbert space ${\cal H}_\gamma$, of dimension higher than that of
${\cal H}_\alpha$ (if that is finite), along with states
$|v_\gamma(b_A,s_k) \rangle$ that are functions not only of Alice's
commands in $A_\gamma = A_\alpha$, but also of the skew $s_k$ of
Bob's clock.  We conclude that for model $\alpha$ to accord with the
behavior of an imperfectly synchronized receiver, there must be an
envelopment $f$ of model $\alpha$ by such a model $\gamma$ with the
property that: \begin{equation} (\forall\ \mbox{$-s$}_0 < s_k < s_0)\ 
f(|v_\gamma(b_A,s_k) \rangle) = |v_\alpha(b_A)\rangle. \end{equation}

An interesting topic for future work is the elaboration of models such
as $\gamma$ and their application to more subtle forms of
synchronization, with which Alice and Bob can advance their cause by
making it more difficult for Eve to make guesses necessary to
eavesdropping.  For example if Alice and Bob manage to synchronize of
Bob's receiver while Alice transmits using a deliberately irregular
clock rate, Eve has a problem determining when to gate her snooping
receiver.

\section{Discussion}

To base QKD security only on proofs is to forget the crucial point
that proofs are purely mathematical.  A proof of a proposition
asserting security shows that the proposition, written as an equation,
follows from assumptions by logic, without guesswork.  Prominent in
the assumptions are inner products.  In simple and rough terms, a
proof of QKD security shows that if Alice's transmitter is modeled by
a certain inner product between two possible states, then the model
says: `any individual eavesdropping attack is detectable.'  But do we
buy the assumption on which the conclusion of the model rests? Can one
decide this by applying logic to experimental tests of the model? More
precisely, suppose relative frequencies of outcomes obtained in a run
of experimental trials agree with the probabilities calculated from
the inner products assumed in the model: do we then buy the model?

We have tried to avoid garbling logical connections within
mathematics, on one hand, with the application of that mathematics to
actual devices on the other.  By analyzing not just a single model but
a multiplicity of models (all purely mathematical) we made plain a
security issue: for any experimental data on hand, more than one model
can fit them perfectly, and among the models that fit the data are
models with widely differing inner products.  Thus choosing one of
these models while rejecting others requires reaching outside the
measured data and the logic of models, as asserted by our Proposition
1. Because inner products, central to QKD security, vary widely over
those models that exactly fit the measured data, the measured data
cannot determine the value of the inner product. Therefore the inner
product is not measurable, and assuming that any model, with its
inner products, applies to particular devices reaches beyond logic
into the realm of guesswork.

Like any proof, our proof of Proposition 1 clarifies the effect of
rules of logic.  This proof escapes the question ``are we sensible to
apply it to cryptography'' because we do not apply it to devices {\em
per se}, but use it only to elucidate the logic of {\em arguments}
about cryptography, arguments that, as it turns out, overlook choices
of model that are apparent once the model that asserts security has
been set in the context of a language game in which it competes with
other models.

The world of cryptography, as we have described it, is a multi-model
world in which each side must guess to choose models that guide
actions. By recognizing feedback, as in section 4, we show choices of
models, with their guesswork, that enter responses to experimental
results, and hence enter the use of outcomes to manage experiments.
Records in the memory of a computer used in a feedback loop can
reflect guesses within a structure of equations and devices.  By
introducing CPCs, with their capacity to mediate between a scientist
and devices and to record guesses as choices of models used in
device-controlling programs, we make a beginning toward exploring the
structure of guesswork in the bridging of equations to devices.

\vskip8pt
{\it Acknowledgments}.  
We thank Howard E. Brandt for reading an early draft and giving us an
astute critique, indispensable to this paper.  We thank
John Lowry, Donald Nicholson, Bahaa Saleh, Alexander Sergienko, Malvin
Teich, and Tai T. Wu for contributing substantially to this paper.
This work was supported in part by the Air force Research Laboratory and 
DARPA under Contract F30602-01-C-0170 with BBN Technologies.

\appendix
\section{Formal definition of QM-models and their
envelopment}\label{app:A}  

To formally present {\em QM-models} and the enveloping of one by
another, it is convenient to introduce a category having structures of
equations of QM-models as its objects \cite{cat}.  We abuse language
by calling these structures QM-models. Any QM-model $\alpha$ consists
of: a domain $C_\alpha$ of commands from CPCs (which we interpret as
commands that set up devices), a set of possible outcomes
$O_\alpha$, a Hilbert space ${\cal H}_\alpha$, and a triple of
functions, namely (1) $|\psi_\alpha \rangle: C_\alpha \times O_\alpha
\rightarrow {\cal H}_\alpha$, (2) $U_\alpha: C_\alpha \times O_\alpha
\rightarrow \{$unitary operators on ${\cal H}_\alpha \}$, and (3)
$M_\alpha:C_\alpha \times O_\alpha \rightarrow \{$semipositive
hermitian operators on ${\cal H}_\alpha \}$, subject to the
constraint that makes $M_\alpha$ work as a positive-operator-valued
measure (POVM), namely that \begin{equation} (\forall\ b \in C_\alpha)\
\sum_{j \in O_\alpha} M(b;j) = \bf{1}, \label{eq:rule}\end{equation}
where the {\bf 1} denotes the identity operator on
${\cal H}_\alpha$.  The triple of functions defines a probability
distribution on outcomes given the set-up $b \in C_\alpha$ by the rule
\begin{equation} \Pr\nolimits_\alpha(j|b) = \langle \psi_\alpha(b;j)|
U^{\dag}_\alpha(b;j) M_\alpha(b;j) U_\alpha(b;j)|\psi_\alpha(b;j)
\rangle. \end{equation} In the examples in the paper, the state
preparation expressed by $|\psi_\alpha \rangle$ depends only on a
command and not on an outcome, as does the unitary transform
$U_\alpha$; in that case equation (\ref{eq:rule}) becomes
\begin{equation}
\Pr\nolimits_\alpha(j|b) = \langle \psi_\alpha(b)|
U^{\dag}_\alpha(b) M_\alpha(b;j) U_\alpha(b)|\psi_{\alpha} (b) \rangle.
\label{eq:simple}\end{equation} We remark that if the devices are
controlled by more than one CPC, as is the case in cryptographic
examples, then the command $b$ that specifies a set-up of devices is
a composite of commands from the various CPCs.  For example if the
set-up is established by the three CPCs of Alice, Bob, and Eve,
respectively, a command $b$ has the form $b = b_A \parallel b_B
\parallel b_E$ where $\parallel$ denotes concatenation of the commands
from the three parties.  

The single set $C_\alpha \times O_\alpha$, which expresses the
possibilities of set-ups of devices for a run of trials along with the
possible outcomes, is the domain for all three functions that
represent state preparation, evolution, and measurement.  Sometimes
one wants to suppose that one device prepares a state, another device
is responsible for evolution, and a third measures.  By adding
assumptions beyond those by which we defined a QM-model, one can
generate a specialization of a QM-model that expresses such a notion,
as if the function $|\psi_\alpha \rangle$ had a domain distinct from
that of $U_\alpha$ and $M_\alpha$.  For example, one can specialize a
QM-model so that the state $|\psi_\alpha(b,j) \rangle$ depends only on
$b_A$ and not on the commands $b_B$ and $b_E$ (concatenated with it to
form $b \in C_\alpha)$, nor on the outcome $j$ in $O_\alpha$.  A
function $|\psi'_\alpha(b_A) \rangle$ can always be written as a
function $|\psi_\alpha \rangle$ on $C_\alpha \times O_\alpha$ for
which variation of the argument $j$ in $O_\alpha$ makes no change in
the value of the function, and similarly, variation in the $b_E$ and
$b_B$ parts of $b$ make no change; for this reason, we can (and do)
look at a function $|\psi'_\alpha(b_E)\rangle$ of one variable as a 
special case of the function $|\psi_\alpha(b,j) \rangle$ of two
variables.  A reason to formulate all three functions with the more
general domain $C_\alpha \times O_\alpha$ is that, as discussed in
\cite{ams}, the supposition that one device prepares a state
independent of other devices is unprovable; it is introducible only as
an additional assumption.  In addition, including the $O_\alpha$
factor in the domain of $|\psi \rangle$ allows the modeling of
randomness in state preparation.

Given two models $\alpha$ and $\beta$ of the category, a {\em
morphism} from $\beta$ to $\alpha$ is defined by a function $f$ from a
subset (not necessarily proper) of $C_\beta \times O_\beta$ to
$C_\alpha \times O_\alpha$, with the property of preserving
probabilities in the sense that: \begin{equation} (\forall\ b \in
C_\alpha, j \in O_\alpha)\ \Pr\nolimits_\alpha(j|b) = \sum_{(b',j') \in
f^{-1}(b,j)} \Pr\nolimits_\beta(j'|b'). \end{equation}

Given such a morphism, QM-model $\beta$ will be said to {\em envelop}
QM-model $\alpha$, and the morphism will be called an {\em
envelopment}.  When the morphism $f$ is such that there are
command-outcome pairs in $C_\beta \times O_\beta$ having positive
probabilities but no image under $f$ in $C_\alpha \times O_\alpha$,
model $\alpha$ is a special case of model $\beta$ with respect to the
probabilities it predicts (but not necessarily with respect to its
internal structure of inner products).

In some cases, an envelopment $f$ from a model $\beta$ to a given model
$\alpha$ preserves inner products in the sense that $|\langle
\psi_\alpha(f(b;j))|\psi_\alpha(f(b';j'))\rangle | = |\langle
\psi_\beta(b;j)|\psi_\beta(b';j') \rangle |$. For most of the
envelopments, however, this is not the case.  It is a mistake in logic
to confuse the property of envelopment with the special case
of envelopment that preserves inner products.

An application of envelopment is the finding, demonstrated in section
\ref{sec:3}, that a model $\beta$ can envelop a model $\alpha$ {\em
without} preserving inner products.  In the case discussed in that
section, there is no `mixing' by $f$, which is to say there exist
functions $g$ and $h$ such that $f(b,j)= (g(b),h(j))$, and the
envelopment of model $\alpha$ by model $\beta$ exhibits the provocative
inequality:
\begin{equation} |\langle \psi_\beta(b)|\psi_\beta(b') \rangle | \neq
|\langle \psi_\alpha(g(b))|\psi_\alpha(g(b')) \rangle| , \end{equation}
which allows for security expressed by a model $\alpha$ to be
contradicted by an enveloping model $\beta$.\looseness=-1  

\section{Security models involving a probe and a defense
function}\label{app:B}

Arguments for the security of quantum key distribution that deal with
noisy channels call for privacy amplification to distill a secure key
\cite{bennett95}.  These arguments center on a {\em defense function}
\cite{slutskyPRA,slutskyAO,brandt2,brandt}.  Defense functions have
been analyzed for models embellished to speak of Eve's use of a probe
\cite{fuchs}.  In such a model $\alpha$, Alice chooses one of several
state vectors in one Hilbert space ${\cal H}_{{\rm sig},\alpha}$ while
Eve generates a fixed vector in a different Hilbert space ${\cal
H}_{{\rm probe},\alpha}$, and the tensor product of Alice's choice of
state vector and Eve's fixed probe vector evolves unitarily in an
interaction, after which Eve and Bob make measurements, Eve confined
to the probe sector and Bob to the signal sector.

To see the consequence of signal leakage for defense functions and
probes, suppose that Alice and Bob use model $\alpha$ which assumes
that Alice chooses between state vectors $|v_\alpha(0) \rangle$ and
$|v_\alpha(1) \rangle$ with inner product having a magnitude $S_\alpha
= |\langle v_\alpha(1)|v_\alpha(0) \rangle |$.  Assuming model
$\alpha$, Alice and Bob determine a defense function $t(n,e_T)$, as
discussed in \cite{slutskyAO}; in order to mark its dependence on
model $\alpha$ and especially its dependence on the inner product of
$S_\alpha$, we write this as $t_\alpha(n,e_T,S_\alpha)$.  As in the
simpler case of section \ref{sec:3}, if the inner products for
distinct signal vectors are all zero, Eve can learn everything without
causing any effect that Alice and Bob can detect; and, as before,
Alice's state vectors are model-dependent, and so are their inner
products.  For this reason, it is easy to adapt the reasoning
of Proposition 1 to prove:

\vskip\abovedisplayskip \noindent {\bf Proposition 2}\hskip1em {\em If
a model $\alpha$ asserts that Alice and Bob can distill a key that is
secure against measurements commanded by Eve from a set of commands
$E_\alpha$, then there exists another model $\beta$ that matches the
predictions of model $\alpha$ for the commands in $E_\alpha$ but, by
virtue of different inner products, makes additional commands
available to Eve that make the key insecure.}

\vfil
\eject
\renewcommand{\thefigure}{\arabic{figure}}
\centerline{\bf Figure Captions}

\vspace{20pt}
\begin{figure}[h]
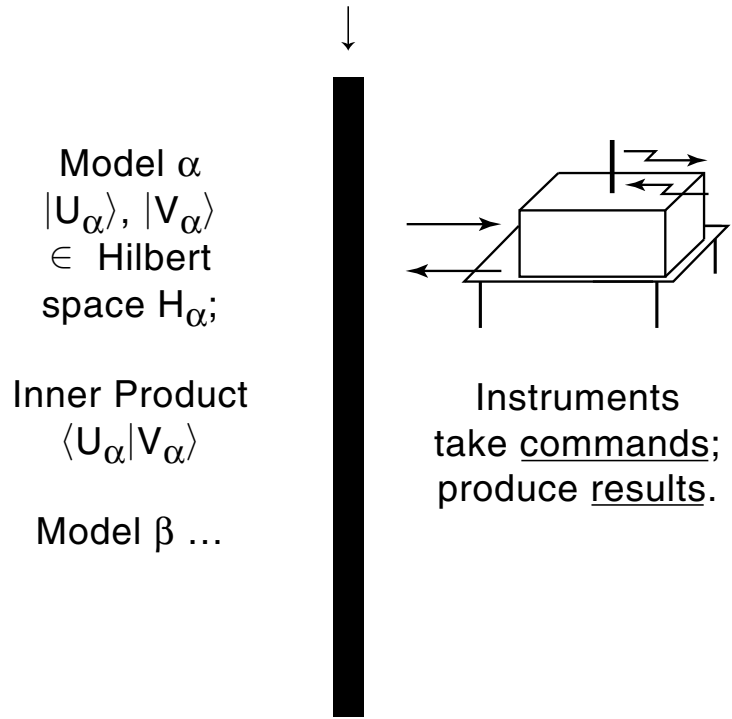
 
\caption{Gap between analyzing with models and using devices}
\label{fig:1}
\end{figure}

\begin{figure}[h]
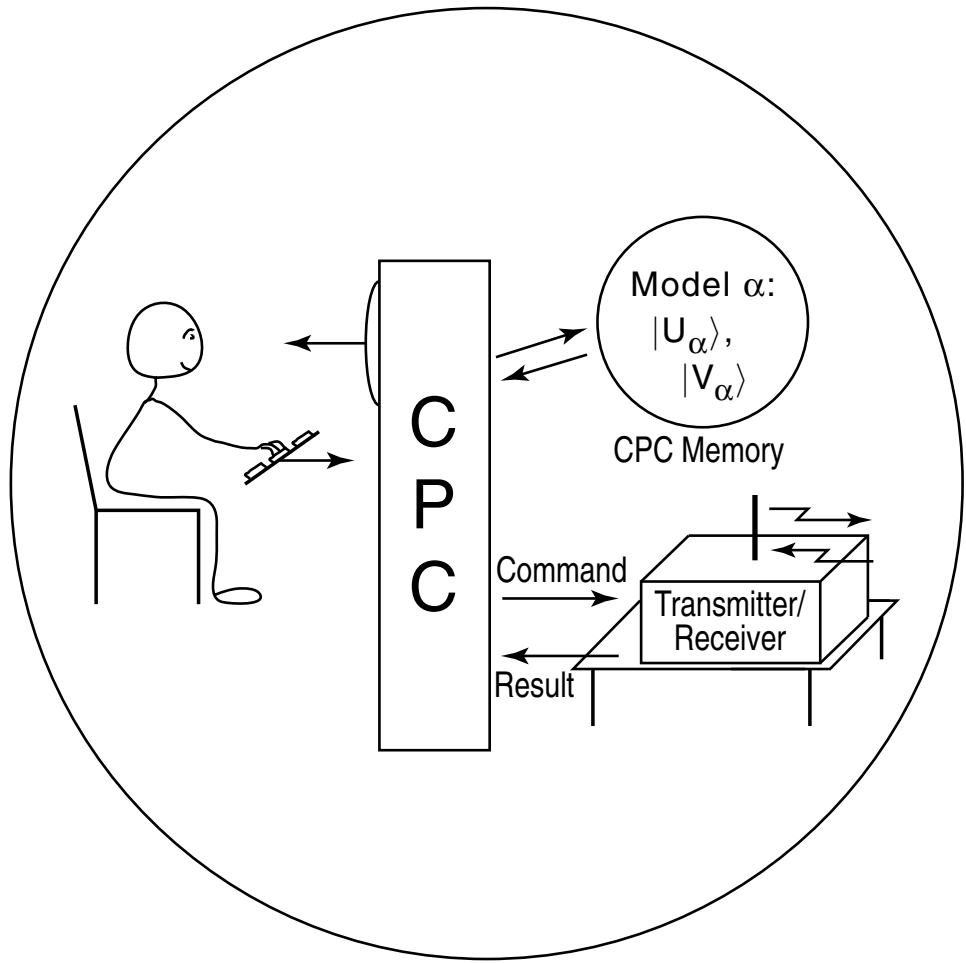
 
\caption{Models and devices reflected in files of CPC}
\label{fig:2}
\end{figure}

\begin{figure}[h] 
\caption{CPC-controlled adjustment in response to outcomes} 
\label{fig:3}
\end{figure} 
\vfil
\eject
\setcounter{figure}{0}
\pagestyle{empty}
\begin{figure}[h]
\vbox to \vsize{\vfil 
\epsfxsize=5in
\centerline{\epsfbox{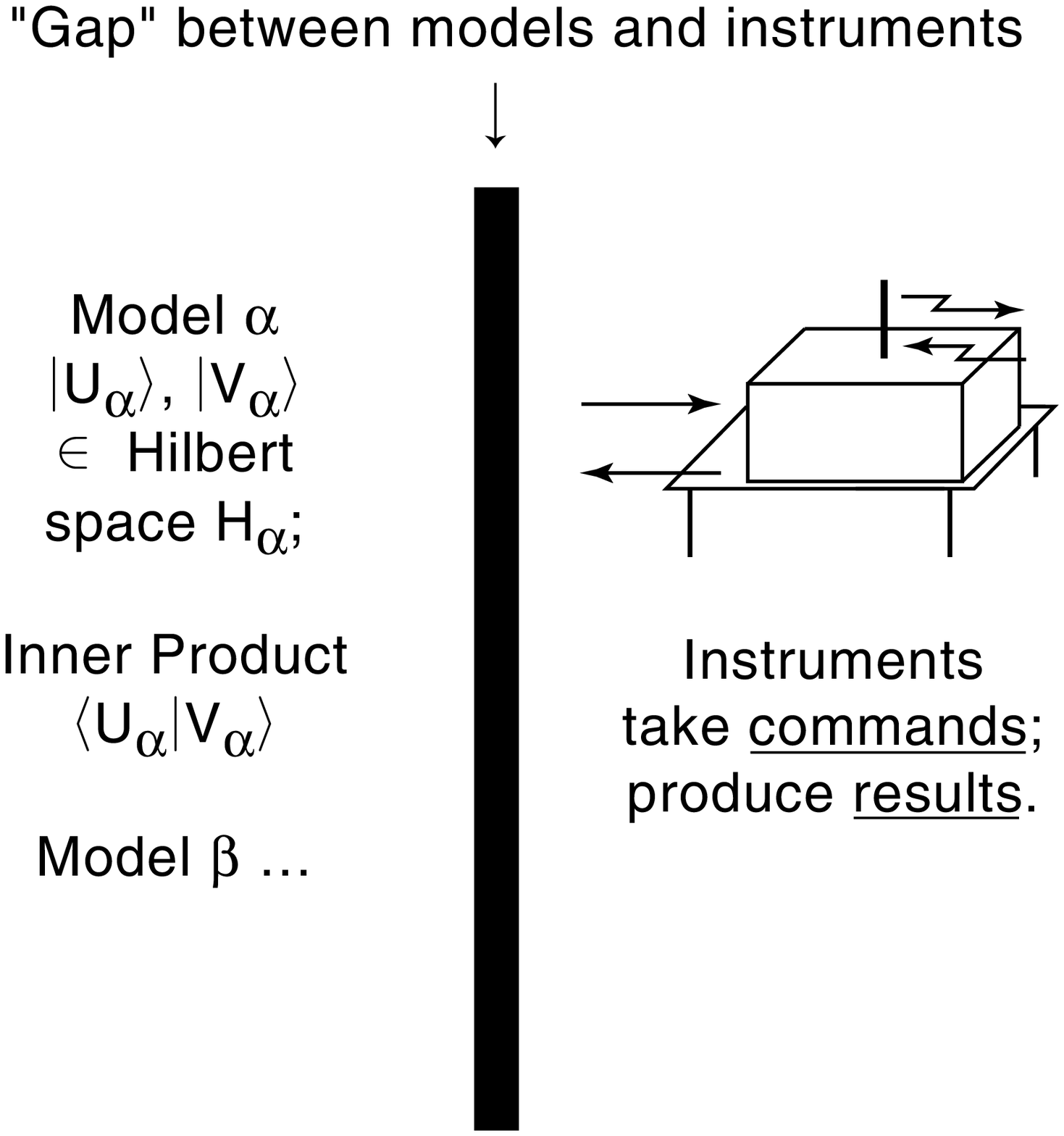}}
\vfil
\caption{\normalsize Gap between analyzing with models and using
devices.}
\vspace{1.25in}}
\end{figure}

\begin{figure}[h]
\vbox to \vsize{\vfil
\epsfxsize=5in
\centerline{\epsfbox{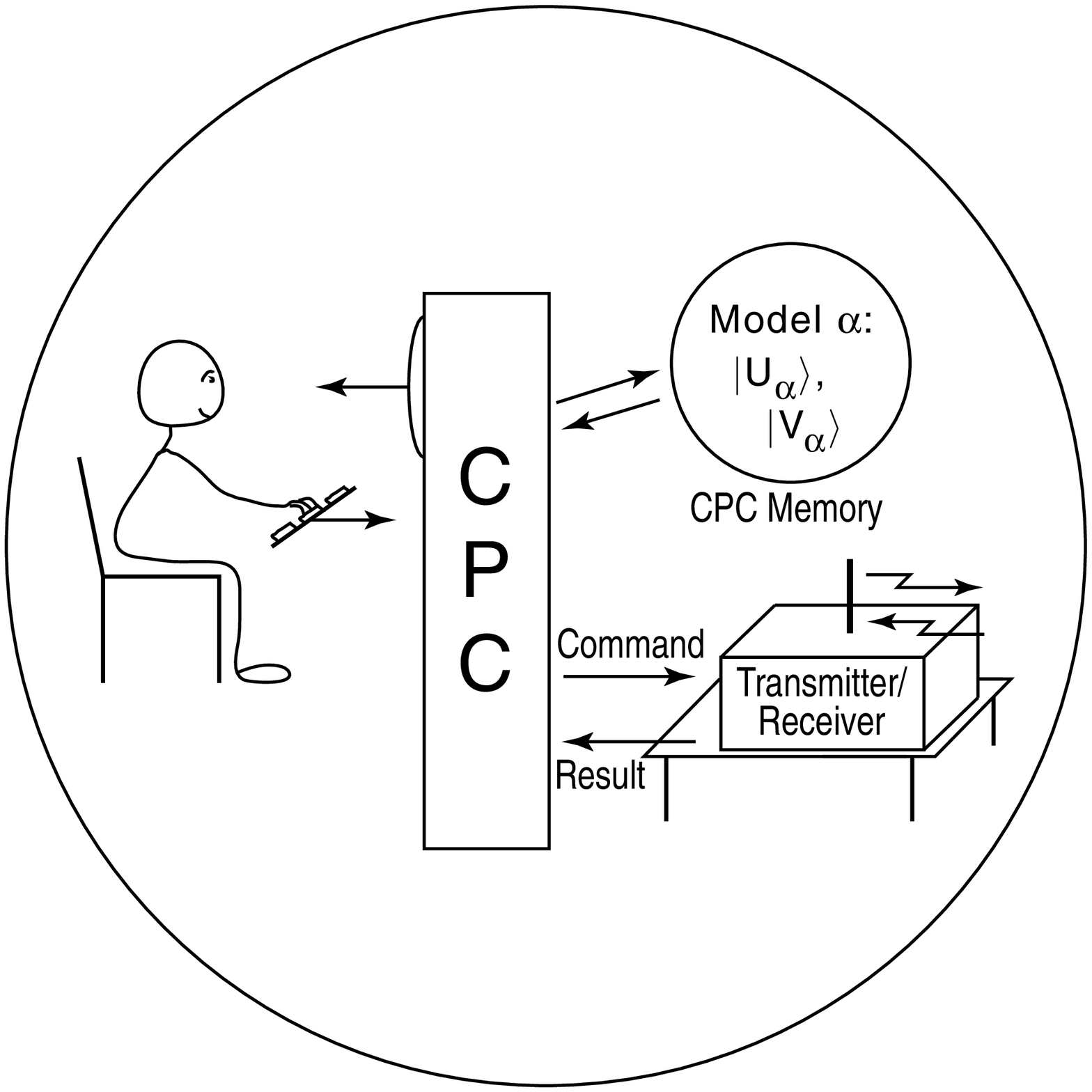}}
\vfil
\caption{\normalsize Models and devices reflected in files of CPC.}
\vspace{1.25in}}
\end{figure}

\begin{figure}[h]
\vbox to \vsize{\vfil
\epsfxsize=5in
\centerline{\epsfbox{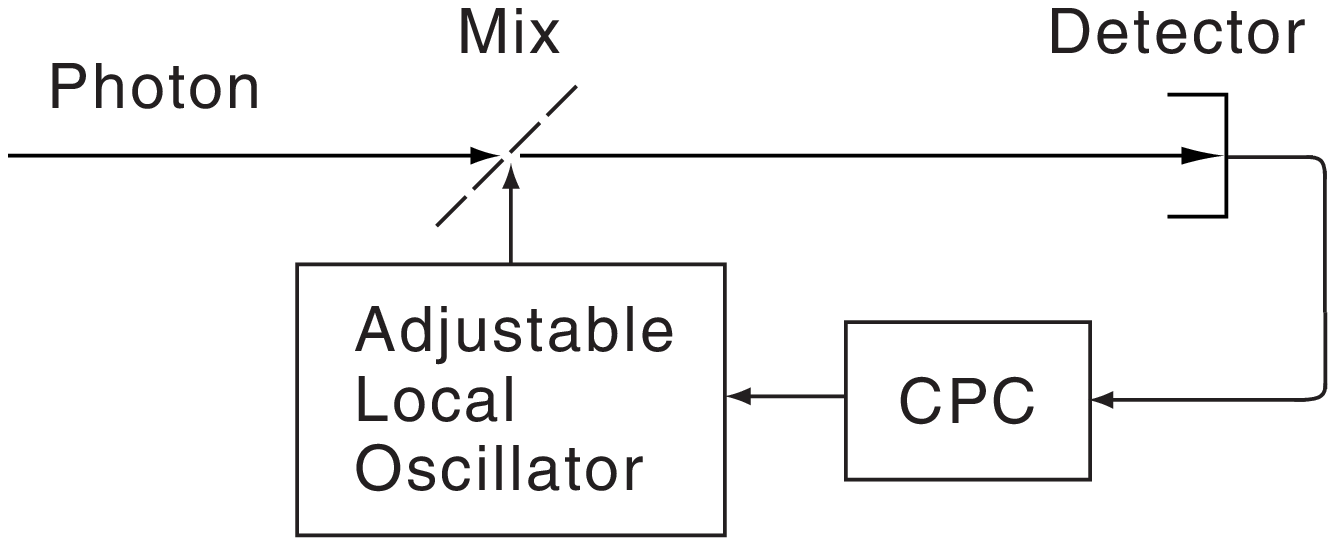}} 
\vspace{40pt} 
\vfil
\caption{\normalsize CPC-controlled adjustment in response to
outcomes.}
\vspace{1.25in}}
\end{figure}
\end{document}